# Origin and Customization of Bandgap in Chiral Phononic Crystals

*Wei Ding[1, 2, †], Rui Zhang[1, †], Tianning Chen[1], Shuai Qu[2,3], Dewen Yu[1, 2], Liwei Dong[2,4], Jian Zhu[1, ]\*, Yaowen Yang[2, ]\*, Badreddine Assouar[5, ]\**

[1]School of Mechanical Engineering and State Key Laboratory of Strength & Vibration of Mechanical Structures, Xi'an Jiaotong University, Xi'an, Shaanxi 710049, P.R. China

[2]School of Civil and Environmental Engineering, Nanyang Technological University, 50 Nanyang Avenue, 639798 Singapore, Singapore

[3]Train and Track Research Institute, State Key Laboratory of Rail Transit Vehicle System, Southwest Jiaotong University, Chengdu, 610031, China

[4]Institute of Rail Transit, Tongji University, Shanghai, 201804, China

[5]Université de Lorraine, CNRS, Institut Jean Lamour, F-54000 Nancy, France

**Abstract**

The wave equation governing the wave propagation in chiral phononic crystals, established through force equilibrium law, conceals the underlying physical information. This has led to a controversy over the bandgap mechanism. In this letter, we theoretically unveil the reason of this controversy, and put forward an alternative approach from wave behavior to formulate the wave equation, offering a new pathway to articulate the bandgap physics directly. We identify the obstacles in coupled acoustic and optic branches to widen and lower the bandgap, and introduce an approach based on spherical hinges to decrease the barriers, for customizing the bandgap frequency and width. Finally, we validate our proposal through numerical simulation and experimental demonstration.

The bandgap property in phononic crystals (PnCs) is associated with extreme spatial dispersion [1], wave guidance [2,3], and thermal physics [4]. Therein, since the inertial amplification effect induced by chirality, which is beneficial for lowering the bandgap beyond the barriers constrained by mass and stiffness [5-7], enables the chiral PnCs the superior performance at low-frequency regime, thus expanding its adaptive scope in the elastic-wave fields [8,9]. However, the bandgap mechanism of chiral PnCs has always been controversial [10-12]. The preliminary theory has indicated the inertial amplification as the mechanism behind such a bandgap [13,14], while different chirality assemblies have different dispersion spectrum [13,15]. Therefore, the mechanism has been attributed to inertial amplification and one mysterious chiral effect of monatomic chains [13].

More recently, two novel explanations have been reported. The first is the dimer chain [12], where coupling longitudinal and torsional waves is similar to the coupled transverse and rotational waves in the periodic mass-spring system [16]. The second explanation is related to analogous Thomson scattering [11], detailing that inertial amplification is induced by coupling two or more polarizations in the same lumped mass and chirality is to achieve the secondary scattering for destructive interferences. These two theoretical interpretations are plausible because of the validation, yet are contradictory since the debate about the existence or absence of inertial amplification.

Here, we develop a theoretical analysis based on the wave behavior in chiral PnCs, to unify and refine the bandgap mechanisms. We demonstrate that the wave equation directly derived from force equilibrium law will conceal the underlying physics, e.g. inertial amplification. Therefore, our method provides another path to articulate bandgap physics and calculate the transmission as well. In contrast to the conventional theoretical method [6,17,18], it allows observing the fundamental physical parameters of acoustic and optical modes under the assumption of elastic ligaments, i.e., inertial amplification coefficient, bending stiffness, stretch stiffness, as well as their origins and interactions. Our analysis pointed out that, the rise of the inertial amplification coefficient is closely related to the bending and stretch stiffness. Therefore, the broad in the bandgap width and the lowering in the starting frequency are restrained by each other, thereby hindering the creation of broad deep-subwavelength bandgaps (the effects of the geometrical dimensions, characterized in equivalent stiffness [19] and equivalent mass [20,21], are considered in normalization). To transcend this barrier, the spherical hinge and a spiral ligament are employed to semi-decouple these coupled physical parameters. The numerical and experimental results validate the correctness of our analysis and the feasibility of our proposal.

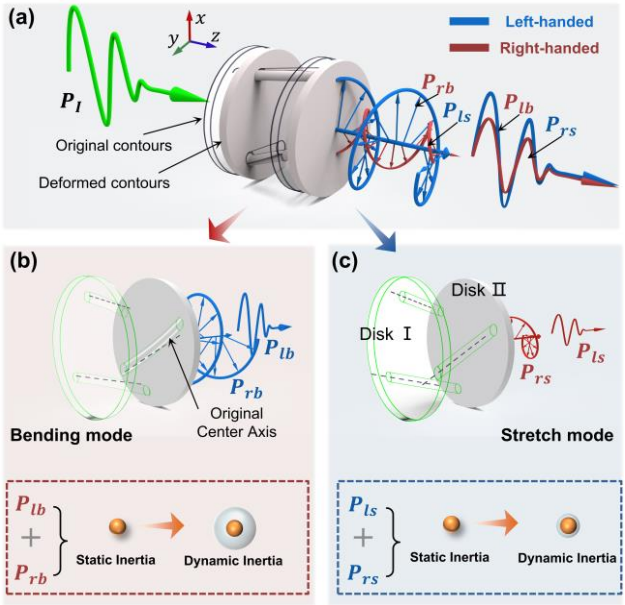

**Figure 1.** (a) Schematics of the chiral subunit cell and its macroscopic deformation under the longitudinal input. (b) Coupled longitudinal and rotational polarizations due to the bending mode of the ligaments. (c) Coupled longitudinal and rotational polarizations due to the longitudinal mode.

In the chiral subunit cell (Figure 1(a)), if there is a longitudinal input $P_I$ on disk1 ($P_I = A_1 e^{-i(wt+\phi_1)}$, where $\phi_1 = 0$), the energy provided by $P_I$ will propagated through the ligaments in the longitudinal and transverse waves simultaneously due to the tilting angle. At deep sub-wavelength scale, we can observe two polarizations at disk II, i.e., longitudinal polarization $P_l$ and rotational polarization $P_r$, which are resulted by the deformation of the tilted ligaments. Therein, the deformation of the ligaments includes two modes, i.e., bending mode (Figure 1(b)) and stretch mode (Figure 1(c)). Notably, the bending mode (Figure 1(b)) is dominated by the transverse waves in the ligaments, whereas the stretch deformation (Figure 1(c)) is dominated by the longitudinal waves.

In the scenario of Figure 1(b), based on the right-hand spiral rule, the disk II will have a longitudinal polarization along $+z$ axis ($P_{lb}$) and a rotation polarization around $+z$ axis ($P_{rb}$) under the bending mode. While in the scenario denoted by Figure 1(c), the disk 2 will have $-z$-axis rotational polarization ($P_{ls}$) in addition to the $+z$-axis longitudinal polarization ($P_{rs}$) under the stretch mode. Notably, the longitudinal polarization $P_{rb}$ and $P_{rs}$ have the same frequency and phase while the rotational polarization $P_{rb}$ and $P_{rs}$ have the same frequency but opposite phase.

Therefore, there are 4 polarizations in disk II, i.e., $P_{lb}$, $P_{rb}$, $P_{ls}$, and $P_{rs}$, where $P_{ij}$ denotes that the $j^{th}$ deformed mode of the ligaments induces the $i^{th}$ polarization of the disk. In detail, subscript $i$ can be longitudinal polarization $l$ or rotational polarization $r$. Subscript $j$ denotes the $j^{th}$ wave mode of the ligaments, which can be bending mode $b$ or stretch mode $s$.

Because $P_{lb}$ and $P_{rb}$ is resulted by the transverse waves of the ligaments, they will have the same frequency and the same phrase at any time. Therefore, for the $i^{th}$ lumped mass, $P_{lb}$ and $P_{rb}$ are linearly correlated and can be determined as

$$U_i^l = u_i^l + \psi_i^l = A_i{}^l e^{-i(wt+\varphi_i^l)} + p(A_{i-1}^l e^{-i(wt+\varphi_{i-1}^l)} - A_i{}^l e^{-i(wt+\varphi_i^l)}) \quad (1)$$

where $p$ denotes the conversion coefficient from longitudinal polarization $P_{lb}$ to rotational polarization $P_{rb}$ and it is characterized as the inertial amplification coefficient in the inertia matrix [11,14].

Like $P_{lb}$ and $P_{rb}$, for the $i^{th}$ lumped mass, $P_{rs}$ and $P_{ls}$ satisfy

$$U_i^s = u_i^s + \psi_i^s = \theta_i{}^s e^{-i(wt+\varphi_i^s)} + q(\theta_{i-1}^s e^{-i(wt+\varphi_{i-1}^s)} - \theta_i{}^s e^{-i(wt+\varphi_i^s)}) \quad (2)$$

where $q$ indicates the conversion coefficient from longitudinal polarization $P_{ls}$ to rotational polarization $P_{rs}$. Because the sense of $q$ is exactly opposite to that of $p$, $q = 1/p$ (see Supplementary S2 for more details).

Because the stretch stiffness $k_s$ is significantly larger than the bending stiffness $k_b$, the wave number of the longitudinal waves is smaller than that of the transverse waves, thus affording phase differences for $P_{rb}$ and $P_{rs}$ ($\varphi^l \neq \varphi^s$).

In the global coordinate system, for the $i^{th}$ lumped mass, the longitudinal displacement $u$ is
$$u_i = u_i^l + u_i^s = A_i^l e^{-i(wt+\varphi_i^l)} + (-1)^i A_i^s e^{-i(wt+\varphi_i^s)} \quad (3)$$
and the rotational displacement $\vartheta$ is
$$\vartheta_i = \psi_i^l - \psi_i^s = \theta_i^l e^{-i(wt+\varphi_i^l)} - (-1)^i \theta_i^s e^{-i(wt+\varphi_i^s)} \quad (4)$$

The potential energy of the system can be divided into $V_b = \sum_{i=1}^{n-1} \frac{1}{2} k_i^b (u_{i+1}^b - u_i^b)^2$ and $V_s = \sum_{i=1}^{n-1} \frac{1}{2} k_i^s (\psi_{i+1}^s - \psi_i^s)^2$ determined by the bending and stretch modes, respectively. The kinetic energy can be divided into $T_l = \frac{1}{2}\sum_{i=1}^{n} m_i \dot{u}_i^2 = \frac{1}{2}\sum_{i=1}^{n} I_i (\dot{A}_i^b + (-1)^i \dot{A}_i^s)^2$, and $T_r = \frac{1}{2}\sum_{i=1}^{n} I_i \dot{\vartheta}_i^2 = \frac{1}{2}\sum_{i=1}^{n} I_i (\dot{\theta}_i^b - (-1)^i \dot{\theta}_i^s)^2$.

If the global translation $u_i$ and rotation $\vartheta_i$ are our concerned variables, the theoretical transmission (Figure 2(b)) and dispersion spectrum (Figure 3 (a)) can be obtained. The final inertia and stiffness matrixes can be written by Eqs. (S15)-(S23), which are similar but not identical to current references like Ref. [12]. In this case, the inertial amplification cannot be observed. The wave equation will only reveal one fact, i.e., the longitudinal polarization is coupled with torsional polarization. However, it has been demonstrated that only specific couplings (such as the syndiotactic PnCs [8,13]) rather than all couplings can give rise to such a bandgap [13,15]. Therefore, the coupling mentioned in Ref. [12] cannot sufficiently elucidate the underlying mechanism of the bandgap.

If regarding $u_i^b$, $u_i^s$, $\psi_i^b$, and $\psi_i^s$ as the concerned variables, we can also obtain the theoretical transmission (Figure 2(b)) which is consistent with numerical results. In this case, several essential information can be captured. First, as denoted by Eqs. (S30)-(S35), the stiffness matrix does not indicate the coupling effect between longitudinal polarization and rotational polarization but the inertial matrix does. Second, the inertial matrix will reveal that the inertial amplification derives from the polarization coupling. Third, Eq. (S29) illustrated that both bending mode and stretch mode can realize the motion coupling and thus obtain the inertial amplification effect, as illustrated by Figure 1(b) and Figure 1(c). Fourth, the motion coupling guided by the bending mode is characterized by longitudinal polarization (because the primary diagonal element of $M_{11}$ includes $m_i$ and the non-diagonal element is $I_i$), while the motion coupling guided by the stretch mode is characterized by the rotational mode (because the primary diagonal element of $M_{22}$ includes $I_i$ and the secondary diagonal element is $m_i$). Fifth, according to Figure 1(b) and Figure 1(c), $P_{rb}$ and $P_{rs}$ have the same frequency but opposite in phase, therefore, there is a degradation between the rotational polarizations determined by bending and stretch modes, which means the larger the dynamic inertia $q$ is, the smaller $p$ is (denoted by Figure 1(b) and Figure 1(c)). Sixth, the bending mode and stretch mode can only determine the range of the inertial amplified-based bandgap and the bandgap will convert into Bragg scattering type after the stretch mode [22].

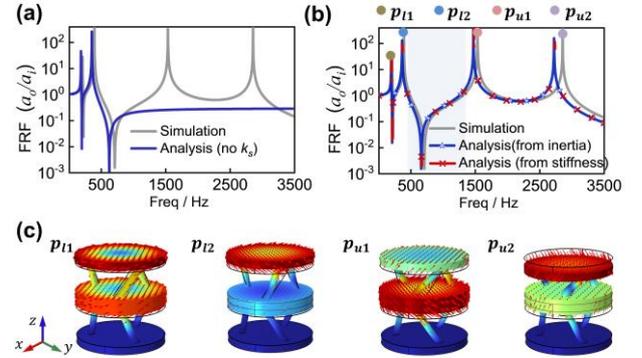

**Figure 2.** (a)-(b) Theoretical and numerical transmissions of the conventional chiral PnCs. "no $k_s$" denotes the results of neglecting the stretch mode; the star-blue line and cross-red lines are the results of considering the stretch mode, where the former one is obtained based on Eqs. S17-S19 and the latter one is obtained based on Eqs. S23-S27. (c) Displacement contours for the upper and lower boundaries of the bandgap.

Notably, Figure 2(b) illustrates that both paths of establishing wave equations can yield identical transmissions to the numerical results. In addition, comparing the deformation contours (shown in Figure 2(c)), the rotational directions of $p_{u1}$ & $p_{u2}$ are opposite to that of $p_{l1}$ & $p_{l2}$ when the translation is along +z axis, which exactly corresponds to the schematics in Figure 1(b) and Figure 1(c), respectively. For convenience, we consider that the lower boundary of the bandgap is the acoustic branch (the two red pass bands in Figure 3(a)) since the vibration in the phase of adjacent atoms, and the upper boundary is the optical branch (the two blue pass bands in Figure 3(a)) because it is similar to that in the long-wavelength limit of an optic mode [23]. The consistency in transmissions Figure 2(b)), dispersion spectrums Figure 3(a) and Figure 3(b)), as well as the deformation schematics Figure 2(b)), can verify the correctness of our analysis.

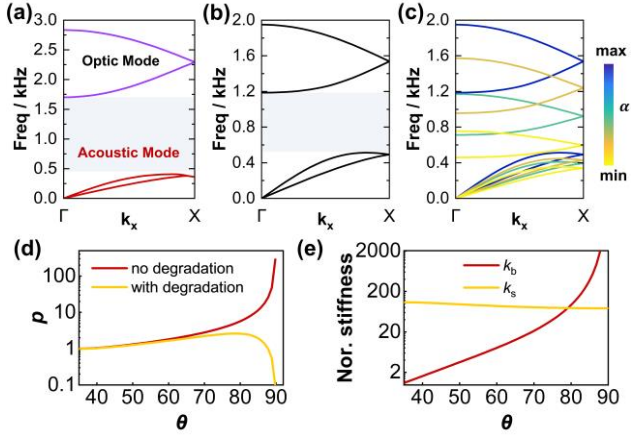

**Figure 3.** (a) Theoretical and (b) numerical dispersion spectrums of the conventional chiral PnCs. (c) Bandgap variation with the different $\theta$ (see Supplementary S5 for details of $\theta$). (d)-(e) Parameter discussion about the influence of $\theta$ on the inertial amplification coefficient $p$, bending stiffness $k_b$, and stretch stiffness $k_s$ of conventional chiral PnCs. The abbreviation "Nor." means normalization of $k/k_r$ where $k_r = 1e6$ N/m (see Supplementary S3 for the governing equation of the dispersion spectrum).

It is crucial to emphasize that these branches essentially differ from conventional diatomic chains. In detail, the upper and lower branches in this context stem from two coupled orthogonal motions that originate from the same atom instead of from two atoms. Coincidentally, this coupled orthogonal polarization introduces a novel control variable for bandgap modulation-namely, inertial amplification [11]. Nevertheless, if we use the traditional theoretical derivation directly based on force equilibrium, the inertial amplification effect will be hidden in the wave equation.

In brief, for chiral PnCs [24,25], it is convenient and pithy to characterize the dispersion spectrum and transmission properties through the wave equation established from force equilibrium, but its final formulas merely present the coupled longitudinal and rotational polarizations, thus obscuring the comprehensive physical insights. Consequently, despite the similarities in coupling orthogonal polarizations observed in other chiral structures [26-28], characterized by auxeticity in quasi-static compression [29], and the systematical governing equations [30-32], there have been limited discoveries of inertial amplification in other chiral PnCs. In particular, the limitation of the wave equation derived from the force equilibrium manifests in the inability to identify significant parameters of the bandgap, such as the inertial amplification coefficient crucial for realizing low-frequency bandgaps [33]. Additionally, it imposes constraints on manipulating bandgaps based on underlying physics and culminates in an evolution primarily through geometric outline [8,10,34].

Furthermore, if neglecting $P_{ls}$ and $P_{rs}$, the theoretical transmission (the green-starred line in Figure 2(a)) is still consistent with the numerical and experiment results in low frequencies [11] and the dynamic equation can also reveal the inertial amplification effect, as demonstrated in Ref. [14]. This is because $k_s$ is usually more than 10 times than $k_b$ [12] and thereby $P_{rs}$ and $P_{ls}$ have the negligible contribution.

The comparison of Figure 2(a) and Figure 2(b) might lead us to believe that the main contribution of stretch mode is to truncate the inertial amplification-based bandgap but it's not true. On the one hand, the bending mode serves to provide the resilience $k_b$ and couple two orthogonal polarizations (rotational and longitudinal polarizations) by bending deformation of the ligaments and thus induce the inertial amplification coefficient $p$. $k_b$ and $p$ are essential for directly determining the acoustic branch. On the other hand, the stretch mode serves to provide the resilience $k_s$ characterized by high stiffness, for the purpose of reducing its own amplification coefficient $q$ to minimize the degradation to $p$. $p$ will be absent if without assistance of stretch mode but the stretch mode will give a degradation to $p$ as well.

In detail, as shown in Figure 3(d), if the degradation is neglected ideally, the amplified dynamic inertia $p$ would be easy to exceed 100 times. However, if considering the degradation, the amplified dynamic inertia $p$ can only be 2.6 times. On the other hand, as illustrated by Figure 3(e), the difference between $k_b$ and $k_s$ will be smaller and smaller with the increase of $\theta$, which is not conducive to achieving a broad bandgap [12]. Ultimately, the upper boundary will approach the lower boundary of the bandgap, leading to the closure of the bandgap, as depicted in Figure 3(c).

In short, due to the coupling between these coupled parameters, it remains a formidable challenge to achieve the objectives of broadening, lowering, and manipulating the bandgap beyond the barriers constrained by mass and stiffness by addressing either of the strategies ($k_b$, $k_s$, and $p$) in isolation.

To decrease the barriers, we propose the strategy as shown in Figure 4(a) to achieve semi-decoupling. Therein, the spiral ligaments provide $k_b$ and the spherical hinges (see Supplementary S4 for the governing equation of the spherical hinge) are responsible for $k_s$ and provide the rotational polarization thus achieving $p$. Regarding the PnC shown in Figure 4(b), because the material component of the spherical hinge is steel, $k_s$ and $k_b$ have great discrepancy, on the one hand, it is benefit to raise the optic branch and thus broaden the bandgap. Meanwhile, the deformation of the stretch mode will be much weaker, so the degradation to $p$ will be weaken. Therefore, the numerical inertial amplification coefficient $p$ can be up to 13 times, as shown in Figure 4(d) (the original coefficient is a maximum of 2.6). Regarding the unit cell, a bandgap extending from 35 Hz-1650 Hz (see Figure 4(c)) can be obtained in the dispersion spectrum (see Supplementary S6 for more details of the simulation). The ratio of the lower boundary of the optic branch to the upper boundary of the acoustic branch is up to 47 times.

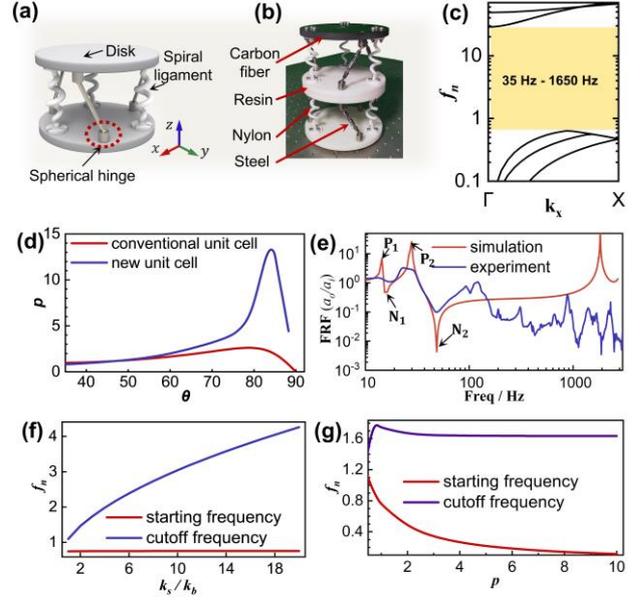

**Figure 4.** (a) Schematics of subunit cell (see Supplementary S5 for details about geometry). (b) Photograph of the experimental sample (See Supplementary S7 for experimental details). (c) Dispersion spectrums of the chiral PnCs. Normalization $f_n = 2\pi f/(\sqrt{k_b/m_1})$, where $m_1 = 0.177$ kg and $k_b = 2.33e4$ N/m. (d) Variation of $p$ with different $\theta$. (e) Numerical and experimental transmission of one unit cell. $P_1$ and $P_2$ denote the resonance peaks while $N_1$ and $N_2$ denote the anti-resonance notches. The green area indicates the bandgap range. (f)-(g) Normalized bandgap width in different stiffness ratios ($k_s/k_p$) and different inertial amplification coefficients $p$. (see Supplementary S6 for details about the simulation)

To verify our proposal under the minimal interference of uncontrollable factors to ensure the validity of the experiment, one unit cell was fabricated and tested. To avoid the local resonance of the lumped masses, the end of the period direction is replaced by a carbon fiber plate, which can provide a high elastic modulus with a low density (see S6 for more details of the experiment). The experimental and numerical results are shown in Figure 4(e). One can see that there is an obvious attenuation after 35 Hz and the experimental and numerical results are in satisfying agreement in 100 Hz, especially at resonance peaks ($P_1$ and $P_2$) and anti-resonance notches ($N_1$ and $N_2$). There are significant deviations between numerical and experimental results after 100 Hz, which might be resulted by the nonlinear collisions from the clearance in the spherical hinge [35].

Because the functions of the spiral spring and spherical hinge are independent, the disparity between $k_s$ and $k_b$ can be magnified by variations in the material and dimensions of the spherical hinges, consequently, the bandgap width can be expanded (Figure 4(f)), where the upper boundary will shift to the higher frequency while the lower boundary is constant.

Besides, $p$ is also independent with $k_b$, which enables to increase $p$ through the tilt angle $\theta$ to decrease the bandgap starting frequency, as illustrated in Figure 4(g). In this case, the lower boundary will shift to a lower frequency while the upper boundary can be constant. While this work showcases realization in broad and low-frequency bandgaps, it should be acknowledged that enhancing the attenuation intensity of the inertial amplification-based bandgap will be the next significant challenge [14].

In summary, in this research, we have theoretically revealed that the inertial amplification effect evolves from inertia matrix to stiffness matrix, and thus unify two ostensibly conflicting explanations of the bandgap mechanism. Based on our theory which allows to observe the comprehensive physics of acoustic and optic branches in chiral PnCs, we have clarified that the close relations between the rise of the inertial amplification coefficient and the bending and stretch stiffness, as well as the restrictions from this close relations on the creation of broad deep-subwavelength bandgaps under boundaries constrained by the constant equivalent density, equivalent stiffness, and lattice constant. Therefore, we have used spherical hinges to achieve the semi-decouple, thus releasing the mutual negative effect between the acoustic and optic boundaries. The numerical and experimental results have confirmed the effectiveness of our proposed scheme and demonstrated that the underlying physics obtained from the wave behavior is instructive for structural design. This work may be able to shield light on the discovery of the inertial amplification effects in other high-dimensional artificial structures, to realize ultra-low-frequency and ultra-broad bandgaps without the requirement of the bulky static mass and fragile static stiffness, as well as to customize the bandgap in chiral PnCs.

This work was financially supported by the National Natural Science Foundation of China (No. 12002258), and the China Postdoctoral Science Foundation (No. 2022M712540). Wei Ding is grateful for the support of the China Scholarship Council (No. 202206280170). Wei Ding appreciates Yuhan Hu (Hohai University) for her guidance in drawing.

* Corresponding authors:
jianzhuxj@xjtu.edu.cn (J. Zhu),
cywyang@ntu.edu.sg (Y. Yang),
badreddine.assouar@univ-lorraine.fr (B. Assouar).
† These authors contributed to the work equally